# Two physical explanations of the nonextensive parameter in a self-gravitating system

Zheng Yahui[1,2] and Du Jiulin[1]

1: *Department of Physics, School of Science, Tianjin University, Tianjin 300072, China*
2: *Department of Physics, School of Science, Qiqihar University, Qiqihar City 161006, China*

**Abstract** In this paper, we present two possible physical explanations of the nonextensive parameter in a self-gravitating system. One is related to the detailed balance in such system. With the detailed balance, the statistical balance of molecular energy is reached, under which the radiation energy "stored" in the molecules becomes important. Then the relation between the nonextensive parameter and the storage coefficient which measures the ability of molecules to store the radiation energy is constructed. The other explanation is from one dimensionless quantity called (logarithmic) temperature gradient defined in the stellar physics. In view of this, we find that in order to keep the nuclear reactions alive inside the Sun, the ratio of radiation pressure to the total pressure in the core must be more than 0.2833.

In recent years, nonextensive statistics mechanics based on Tsallis entropy [1] has been applied to many multidisciplinary fields, remarkably such as plasmas [2] and astrophysics [3]. One important problem remained until now is that the physical essence of the entropy index or the nonextensive parameter *q* is still under discussion. However, what is verified is that in different system this parameter has different origin. In the systems with long-range interactions, such as self-gravitating systems and plasmas, the nonextesnive parameter is mainly related to the long-range potentials and electromagnetic fields [4]. In a general stochastic dynamical system, the parameter may be originated from the noise and friction as well as the associated fluctuation-dissipation relation [5].

In this work, we present two physical explanations of the origins of the nonextensive parameter in the background of classical statistics. One is related to the detailed balance in a self-gravitating system. The other is originated from the dimensionless quantity, called (logarithmic) temperature gradient, defined in the stellar physics.

Before this, let us firstly analyze the concept of temperature in the classical statistics. Consider one classical gaseous system, where the interactions between the molecules in the system is short range, and there is no any external field applied to the molecules. When this system is isolated, it can approach the thermal equilibrium, one obvious character of which is that there exists not any dissipative process in the whole system. In order to have an insight into this, we divide this system into two parts A and B by a imaginary plane, and then consider the exchange of the molecules between them.

We confirm that the whole system is also at the dynamical equilibrium. So the number of molecules transferring from A to B is identical to the one transferring from B to A. Now that the whole system satisfies the Boltzmann-Gibbs distribution, as long as the time is long enough, the transferring molecules can also satisfy the same Boltzmann-Gibbs distribution. Therefore, setting the system temperature as $T$ and taking the system to consists of single-atom molecules, the average kinetic energy of every transferring molecule is $3k_BT/2$.

With the equilibrium condition, the average kinetic energy of transferring molecules from A to B must be equal to the one from B to A. Now that the transferring molecules satisfy the same Boltzmann-Gibbs distribution as the non-transferring molecules in every part, the average kinetic energies of molecules inside these two parts are identical to each other, that is,

$$\frac{3}{2}k_BT_A = \frac{3}{2}k_BT_B \tag{1}$$

Now we put the system in the gravitational field of the earth. We assume the energy transfer is along the horizontal direction, so the gravitational potential can be regarded as constant. Then, the energy of the transferring molecule contains the potential. With the same process, we have

$$\frac{3}{2}k_BT_A + m\varphi_{A0} = \frac{3}{2}k_BT_B + m\varphi_{B0} \tag{2}$$

where $\varphi_0$ is the constant potential. The equations (1) and (2) show that, when the system is at the thermal equilibrium, the average energy of molecules balances everywhere in the whole system. Here, the average is obtained in the local sense, so it is possibly different in different parts of the system. Therefore, equations (1) and (2) can be called statistical balance of molecular energy. Obviously, from these two equations one obtains

$$T_A = T_B \tag{3}$$

This is the natural result for the classical system at thermal equilibrium. However, for the self-gravitating particle system, the result is not so concise.

In the self-gravitating system, the long range gravitational interaction is always treated as one mean field. It is obvious that the self-gravitational potential is inhomogeneous in the whole system, and at the same time the temperature is also inhomogeneous. For simplicity, we assume the system is spherically symmetric. In every spherical shell, we adopt the local equilibrium assumption; so the molecules in the shell follow the classical Boltzmann-Gibbs distribution.

When the self-gravitating system is at the hydrostatic equilibrium, between any two adjacent spherical shells ($A_i$ and $A_j$, $i, j=1,2,…$) the numbers of exchange molecules are equal to each other. When a molecule transfers from one shell to another shell, the mean field can do some work on this molecule, and now that these two shells have different (Lagrange) temperature, its average kinetic energy also changes. Therefore, as long as there is not any energy dissipation in the whole self-gravitating system, there must be

$$\frac{3}{2}k_BT_i - \frac{3}{2}k_BT_j = -m(\varphi_i - \varphi_j), \quad i,j = 1,2,\cdots \tag{4}$$

or

$$\frac{3}{2}k_B T_i + m\varphi_i = \frac{3}{2}k_B T_j + m\varphi_j, \quad i,j = 1,2,\cdots \tag{5}$$

This is the statistical balance of molecular energy. On the other hand, (5) indicates

$$\frac{3}{2}k_B T + m\varphi = const \tag{6}$$

The gradient of the above expression is

$$\frac{3}{2}k_B \nabla T + m\nabla \varphi = 0 \tag{7}$$

In nonextensive statistics mechanics, it is supposed that the microscopic dynamical behavior of a self-gravitating particle (single-atom molecule) system at hydrostatic equilibrium is governed by the generalized Boltzmann equation [6]

$$\frac{\partial f_q}{\partial t} + v \cdot \frac{\partial f_q}{\partial r} - \nabla \varphi \cdot \frac{\partial f_q}{\partial v} = C_q(f_q) \quad, \tag{8}$$

where $\varphi$ is the self-gravitational potential determined by the Poisson equation, and $C_q$ is the $q$-collision term. According to the $q$-H theorem, when the collision term vanishes in (8), the distribution evolves irreversibly towards the generalized Maxwellian $q$-velocity distribution [6,7]

$$f_q(r,v) = nB_q \left(\frac{m}{2\pi k_B T}\right)^{\frac{3}{2}} \left[1 - (1-q)\frac{mv^2}{2k_B T}\right]^{\frac{1}{1-q}}, \tag{9}$$

where $n$ is the number density of particles, $B_q$ is the q-dependent normalized constant. From this distribution function (9), one important relation was obtained [8]

$$k_B \nabla T + Qm\varphi = 0, \tag{10}$$

with the condition

$$\nabla Q \equiv \nabla(1-q) = 0 \quad. \tag{11}$$

The equation (10) is actually the natural result of detailed balance. Let the right hand side of (8) be zero, substitute (9) into (8), and compare the coefficients of same order power of the velocity in both sides of (8), one can easily get (10).

Compare (10) with (7), one can find (7) is the special case $Q=2/3$ of (10). This is not satisfactory now that the self-gravitating system at hydrostatic equilibrium yields to distribution (9) and the parameter $q$ is not assigned one special value. Actually, for such a self-gravitating system where the gradient of temperature is not zero, the energy transfer caused by the thermal radiation should be considered. Therefore, we add one radiation term into (6), that is,

$$\frac{3}{2}k_B T + \lambda k_B T + m\varphi = E_C, \tag{12}$$

where the $E_C$ is a constant in the whole system. The radiation term denotes the average radiation energy stored in every molecule.

Here, we think this radiation term is proportional to the Lagrange temperature $T$, which has two sources. One is the average energy of the electrons ionized from the molecules (due to the absorption of radiation energy), assumed to follow the Boltzmann-Gibbs distribution in a self-gravitating system. So its average energy is proportional to $T$. The other source is the radiation field in the spaces between the molecules. The radiation field is resulted from the scattering of ionized electrons to the photons and the transition of bound electrons between different energy level in the molecules. According to the quantum statistics, the total energy of the radiation field in equilibrium is proportional $T^4$ to and the total (average) photon number in a given field is proportional to $T^3$; therefore, the average energy of every photon is proportional to $T$. Although there are two different sources, we regard the radiation energy as being 'stored' inside the molecules and it can be transfered with the molecules.

In order to clarify the physical meaning of (12), let us consider the Fig.1. When the whole system is at the hydrostatic equilibrium, the numbers of exchange molecules between the adjacent spherical shells A and B are equal to each other. The quantity ε denotes the sum of the average molecular kinetic energy and the average radiation energy 'stored' in the molecule. If the sum of ε in each shell and its corresponding potential is a constant $E_c$, ordinarily called statistical balance of molecular energy, now that we have assumed each shell is at the Boltzmann thermal equilibrium, there must not be any energy dissipation in the whole system. The entropy balance, and then the detailed balance hold in the self-gravitating system with the inhomogeneous temperature. Then the system yields to the generalized Maxwellian distribution (9). The detailed balance implies the whole system approaches to one "equilibrium" different from the classical Boltzmann thermal equilibrium. Generally, we can call this Tsallis equilibrium.

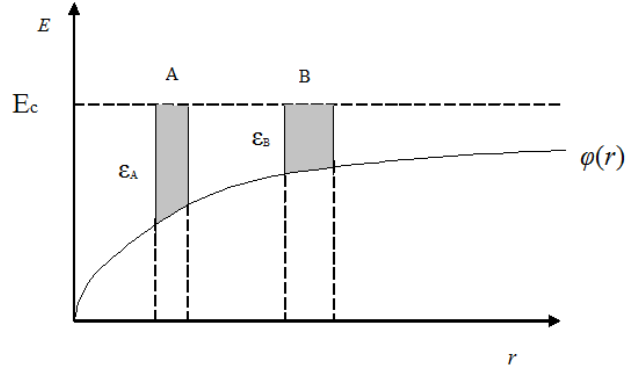

Fig 1. The schematic diagram of detailed balance, assuming the self-gravitating system is spherically symmetric. Under this condition, the whole system approaches the Tsallis equilibrium. The quantity ε denotes the sum of the average molecular kinetic energy and the average radiation energy stored in the molecule. The potential changes along the radial direction.

With the statistical balance of molecular energy, calculating the gradient of (12), one has

$$k_B \nabla T + \frac{1}{3/2 + \lambda} m \nabla \varphi = 0. \tag{13}$$

Obviously, if defining

$$Q = \frac{1}{\frac{3}{2} + \lambda}, \tag{14}$$

equation (13) is completely identical to equation (10). The parameter $\lambda$ can be called the storage coefficient, which measures the ability of the molecules 'to store' the radiation energy. So, based on the classical statistics, one can construct the relation (10) which is derived from the nonextensive statistical theory. When the statistical balance of molecular energy is satisfied, now that we have assumed that each shell is at the thermal (Boltzmann) equilibrium (so there is no energy flux inside the shell), there must not be any energy dissipation in the whole system. Therefore, the relation (10) is equivalent to the detailed balance principle.

In the process deriving equation (13) we apply the local equilibrium assumption, so the Boltzmann-Gibbs distribution is suitable for the local sense. On the other hand, the whole system at the hydrostatic equilibrium yields to the generalized Boltzmann-Gibbs distribution at Tsallis equilibrium [9]. These two distribution functions are both suitable in the self-gravitating system. One is correct in the local sense, and the other is valid in the whole sense. This coexistence of distributions indicates the assumption of temperature duality [10], which states that the Lagrange temperature related to the local thermal (Boltzmann) equilibrium and the physical temperature [11] related to the Tsallis equilibrium in the whole sense are both necessary in the self-gravitating system.

With the assumption of temperature duality, one can introduce a new concept of the temperature in the self-gravitating system. When the whole system is at the Tsallis equilibrium, or the statistical balance of molecular energy is satisfied, equation (10) or (13) holds. For describing the 'equilibrium' state, one generalized temperature as the state parameter is required. This can be done through integrating (10) with the condition (11). Then we obtain

$$T_g = T + \frac{Qm\varphi}{k_B} + T_0 \qquad (15)$$

Apparently, the definition includes the gravitational potential, so ordinarily we call it gravitational temperature. If set $T_0=0$, compare (15) to (12) one can easily find that they are identical to each other. That is, equation (12) actually gives the definition of gravitational temperature in the classical statistics background.

Fig. 1 and equation (13) present one physical explanation of nonextensive parameter $Q=1-q$ in the framework of classical Boltzmann statistics. From the expression (13) it is easy to find that as long as the inhomogeneous potential exists the parameter $Q$ can not be zero. So the velocity distribution function must be the power law one. Because possible values of the storage coefficient is $0< \lambda <\infty$, this physical explanation in eq.(14) offers one powerful proof of $0< Q <2/3$ if the system consists of single-atom molecules..

Furthermore, if the statistical balance of molecular energy were violent, the equation (12) could not hold in the whole system. However, we can apply the generalized local (Tsallis) equilibrium assumption (based on which the generalized heat flux can be defined [12]), which states that, in the scale ($L$) far more than the one (the thickness $l$) of the spherical shell in Fig 1, that is $L \gg l$, the generalized Maxwellian distribution (9) holds. So in this scale $L$, the quantity $E_C$ in (12) is still a constant and the gravitational temperature can be also defined. Anyway, the expression (14) holds in general case.

With the solar standard date BS2005, the storage coefficient can be calculated according to (14), since the nonextensive parameter $Q$ could be obtained through relation (10). The distribution curve of this storage coefficient along the radial direction is drawn in Fig 2. One can easily find that, the coefficient is about unity in the convective zone and its maximum value appears in the radiation region, but not in the core region. Obviously, the value range is limited, showing the limited ability of molecules to store radiation energy. According to the analysis of the radiation term in (12), the storage coefficient is related to the degree of ionization of molecules and the average number of photons interacting with one ionized or bound electron. It is apparent that the limitation of its value is originated from the limitations of degree of ionization and average number of photons. Furthermore, the storage coefficient can be linked to the difference of adsorption coefficient and emission coefficient in the radiative transfer process. So from the curve of Fig.2, one can find some information about the adsorption coefficient (opacity) and emission coefficient in the solar interior.

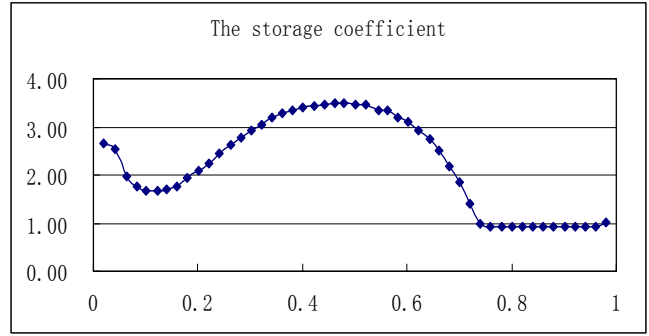

Fig 2. The distribution of storage coefficient along the radial direction inside the solar interior.
It reveals the change of ability of molecules to store the radiation energy with the position.

Of course, apart from the physical explanation of the parameter $Q$ with (13), we have another explanation about this parameter. In the stellar physics, one dimensionless quantity called logarithmic temperature gradient (to distinguish it from the ordinary temperature gradient, we call it logarithmic temperature gradient) is defined as follows [13],

$$\hat{\nabla} \equiv \frac{d \ln T}{d \ln P}. \tag{16}$$

One self-gravitating system at hydrostatic equilibrium satisfies the equation,

$$\nabla P = -mn \nabla \varphi. \tag{17}$$

Considering the spherically symmetry, one can obtain

$$\frac{dT}{dr} = \frac{dT}{dP}\frac{dP}{dr} = -mng\frac{T}{P}\hat{\nabla}. \tag{18}$$

The quantity $g$ is the gravitational acceleration. If regarding the system as one ideal gas sphere with the pressure $P = nk_B T$, one has

$$\frac{dT}{dr} = -\frac{mg}{k_B}\hat{\nabla}. \tag{19}$$

Compare the equation to the scalar form of (10), one immediately obtains

$$Q = \hat{\nabla}. \tag{20}$$

That is to say, if the pressure of the system only includes the gas pressure, the nonextensive parameter Q is just the logarithmic temperature gradient (16). Again, this quantity can not be zero under the condition considered.

Ordinarily, the pressure of system includes the radiation pressure apart from the gas pressure. So $P=P_g+P_r$, where $P_g$ is the gas pressure, and $P_r$ is the radiation pressure. For convenience, we define the quantity

$$\beta \equiv P_g / P. \tag{21}$$

With the gas pressure $P_g=nk_BT$, (18) becomes

$$\frac{dT}{dr} = -mg\frac{\beta}{k}\hat{\nabla}. \tag{22}$$

So one has

$$Q = \beta\hat{\nabla}. \tag{23}$$

Inside the star, the convection instability criterion is generally expressed as [13]

$$\hat{\nabla} \geq \hat{\nabla}_{ad}. \tag{24}$$

Assume the single-atom ideal gas inside the star is completely ionized, one has [13]

$$\hat{\nabla}_{ad} = \frac{1+(1-\beta)(4+\beta)/\beta^2}{5/2+4(1-\beta)(4+\beta)/\beta^2}. \tag{25}$$

Then inequality (24) becomes

$$Q \geq \beta\hat{\nabla}_{ad} = \frac{\beta+(1-\beta)(4+\beta)/\beta}{5/2+4(1-\beta)(4+\beta)/\beta^2}. \tag{26}$$

In the convective layer, ordinarily there is $\beta \to 1$, so one has

$$Q \geq 2/5. \tag{27}$$

This is the convective instability criterion near the star surface [14], expressed by the nonextensive parameter. The instability criterion is important to analyzing the convection near the stellar surface. At the stellar center the nuclear reaction requires the existence of convection, so the inequality (26) also holds. Here, the radiation pressure plays an important role. Define the following function

$$f(\beta) = \frac{\beta+(1-\beta)(4+\beta)/\beta}{5/2+4(1-\beta)(4+\beta)/\beta^2} = \frac{4\beta-3\beta^2}{16-12\beta-3\beta^2/2}, \tag{28}$$

one can prove that this function increases monotonously with $\beta$ for $0 < \beta < 1$.

Setting the value of $Q$ at the stellar center as $Q_c$, from (26) we can calculate the critical value of parameter $\beta$, that is,

$$\beta_c = \frac{12Q_c + 4 - 4\sqrt{15Q_c^2 - 6Q_c + 1}}{6 - 3Q_c}. \tag{29}$$

Then the convection instability criterion at stellar center is written as

$$\beta \leq \beta_c. \tag{30}$$

This means that, in order to keep the convection alive at the stellar center, the gas pressure must be small enough.

With the data of solar standard model BP2000, one can estimate the value of $Q_c$. We roughly adopt the following formula

$$Q_c = \frac{k_B(1.568 \times 10^7 - T_r)}{(r - 0.0065 R_\odot) m_r g_r}. \tag{31}$$

And we adopt $r=0.042 R_\odot$, where $R_\odot$ is the solar radius. At this position, the average molecular mass, gravitational acceleration and temperature is respectively $m_r$, $g_r$, $T_r$, whose value can be obtained from the standard model. Then one obtains $Q_c \approx 0.2$. Therefore the critical value $\beta_c=0.7167$. That is to say, in order to keep the nuclear reaction alive in the solar core, the ratio of radiation pressure to the total pressure must be more than 0.2833.

In summary, we present in this paper two explanations to the origins of nonextensive parameter $Q$ in the self-gravitating gaseous system, based on the classical statistics theory. One is related to the detailed balance, under which condition the molecular energy can reach the statistical balance. So the average of total energy of molecules in different parts of system is a constant. Here, the average is made under the local (Boltzmann) thermal equilibrium condition. With the statistical balance of molecular energy (12), the expression relating the parameter $Q$ to the storage coefficient in (14), which measures the ability of molecules to store the radiation energy, is also obtained.

When the detailed balance and then the statistical balance of molecular energy (13) are satisfied, the whole self-gravitating system is at an 'equilibrium' state with inhomogeneous (Lagrange) temperature, which is apparently different from the classical thermal equilibrium concept. So we call it Tsallis equilibrium, which is described by the generalized Maxwellian velocity distribution (9). Therefore in a self-gravitating system there are two different distributions. One is the Boltzmann-Gibbs distribution which is suitable for the local sense, and the other is the nonextensive distribution (9) in the whole sense.

The coexistence of two different distributions indicates the assumption of temperature duality, which states two different temperature concepts in the system are both significative. One is the classical (Lagrange) temperature defined based on the local (Boltzmann) thermal equilibrium, thus related to the Boltzmann-Gibbs distribution. The other is the physical temperature based on the Tsallis equilibrium described by the Tsallis power-law distribution, that is, the generalized Maxwellian distribution (9). The physical temperature, called gravitational temperature in the self-gravitating system, is naturally deduced from (12).

In our opinion, the radiation term in (12) is proportional to Lagrange temperature $T$, which is important for our discussion. Just as we mentioned, this term at least has two sources. One is the average kinetic energy of electrons ionized from molecules (due to the absorption of radiation energy), which are assumed to obey to the Boltzmann distribution. So the average electron energy is proportional to $T$. The other source is the radiation field bound to the space between molecules due to the scattering of photons and ionized electrons, and also due to the transition of bound electrons between different energy level in the molecules. According to the quantum statistics, the average energy of photons existing in the radiation field is also proportional to $T$. The average energy of photons can change with the temperature $T$ and also with the average kinetic energy of molecules. So the radiation energy is actually 'stored' inside the molecules. The proportional coefficient in the radiation term then measures the ability of molecules to store the radiation energy, called storage coefficient.

The other physical explanation of the parameter $Q$ is originated from the dimensionless temperature gradient (16) defined in the stellar physics. Assume the ideal gas in the system is completely ionized; the convective instability condition can be expressed by the parameter $Q$ (26), where the ratio $\beta$ of gas pressure to total pressure is defined. From the instability condition, one can find that, in order to keep the nuclear reaction in the sun, the ratio of radiation pressure to total in the solar core must be more than 0.2833. Therefore, the radiation pressure is important in the solar core. We know that the radiation pressure is related to $T^4$, and gas pressure is related to $T$. So when the radiation pressure ratio is large enough, the change of total pressure is sufficiently/more sensitive to the change of Lagrange temperature $T$ in the solar core. Then the nuclear reaction can keep stable easily due to the negative specific heat in this region.

These two explanations of the nonextensive parameter proposed in this paper are both related to the heat radiation inside the self-gravitating system. The first explanation is related to the energy balance (12), and the parameter $Q$ is dependent on the radiation energy term. The second one is related to the dynamic balance (17), and $Q$ is dependent on the radiation pressure to some extent. So they are thermodynamic and dynamic explanations respectively. This provides more insights about the essence of the nonextensive parameter $Q$.

∗ ∗ ∗

This work is supported by the National Natural Science Foundation of China under Grant No.11175128 and the Higher School Specialized Research Fund for Doctoral Program under grant No.20110032110058. Zheng Yahui is also supported by the National Natural Science Foundation of China under Grant No.11405092 and the Heilongliang Province Education Department Science and Technology Research Project No. 12541883.

**References**
[1] C. Tsallis, J. Stat. Phys. **52** (1988) 479.
[2] Lima J. S. A., Silva R. and Santos J., Phys. Rev. E, **61** (2000) 3260;
    Leubner M. P., Astrophys. J., **604** (2004) 469;
    Du J.L., Phys. Lett. A, **329** (2004) 262;
    Liu B. and Goree J., Phys. Rev. Lett. **100** (2008) 055003;
    Liu L. and Du J. L., Physica A, **387** (2008) 4821;
    Liu Z., Liu L. and Du J. L., Phys. Plasmas, **16** (2009) 072111;
    Tribeche M. and Pakzad H. R., Astrophys. Space Sci., **339** (2012) 237;
    Gong J.Y, Liu Z. P. and Du J. L., Phys. Plasmas, **19** (2012) 083706;
    Yu H.N. and Du J. L., Annals Phys. **350** (2014) 302 and the references therein.
[3] Leubner M.P., Astrophys. J., **632** (2005) L1;
    Leubner M. P. and Voros Z., Astrophys. J., **618** (2005) 547;
    Du J.L., Europhys. Lett., **75** (2006) 861;
    Du J.L., Astrophys. Space Sci., **312** (2007) 47 and the references therein;
    Cardone V. F., Leubner M. P. and Del Popolo A., Mon. Not. R. Astron. Soc., **414** (2011) 2265.
[4] Du J.L., Astrophys. Space Sci., **312** (2007) 47;
    Yu H.N. and Du J.L., Annals Phys. **350** (2014) 302.
[5] Du J. L., J. Stat. Mech. (2012) P02006;
    Guo R. and Du J. L., J. Stat. Mech. (2013) P02015.
[6] Lima J. A. S., Silva R. and Plastino A. R., Phys. Rev. Lett., **86** (2001) 2938.

[7] Silva R., Plastino A. R. and Lima J. A. S., Phys. Lett. A, **249** (1998) 401.

[8] Du J.L., Europhys. Lett. **67** (2004) 893.

[9] Du J.L., CEJP, **3** (2005) 376.

[10] Zheng Y. H., Physica A, **392** (2013) 2487-2491.

[11] S.Abe, S.Martínez, F.Pennini and A. Plastino, Phys. Lett. A **281** (2001) 126.

[12] Zheng Y. H. and Du J. L., EPL, **105** (2014) 54002

[13] Huang R. Q., Stellar Physics, first edition, Springer-Verlag, Singapore 1998.

[14] Zheng Y. H., EPL, **101** (2013) 29002.